# Revealing buried information: Statistical processing techniques for ultracold gas image analysis

Stephen R. Segal,[1,*] Quentin Diot,[1,†] Eric A. Cornell,[1] Alex A. Zozulya,[2] and Dana Z. Anderson[1,‡]

[1]*Department of Physics, University of Colorado, and JILA, National Institute of Standards and Technology and University of Colorado, Boulder CO 80309-0440, USA*

[2]*Department of Physics, Worcester Polytechnic Institute, 100 Institute Rd, Worcester MA 01609, USA*

The techniques of principal and independent component analysis are applied to images of ultracold atoms. As an illustrative example, we present the use of these model-independent methods to rapidly determine the differential phase of a BEC interferometer from large sets of images of interference patterns. These techniques have been useful in the calibration of the experiment and in the investigation of phase randomization. The details of the algorithms are provided.

---

[*] segals@colorado.edu

[†] Current address: Department of Radiation Oncology, University of Colorado Denver, Mail stop F-706, 1665 Aurora Ct Ste 1032, Aurora CO 80045, USA

[‡] dana@jila.colorado.edu

# I. INTRODUCTION

Fits of images to theory models have played a central role in the measurement and analysis of ultracold gases. An absorption or phase-contrast image of a cloud of atoms depicts the instantaneous density distribution. From one image (or sometimes many images), one typically wishes to extract just a few parameters of interest from the thousands to millions of pixel values. In most experiments, this parameter extraction task is accomplished by fitting the imaged distribution to a model. This model usually is derived from *a priori* knowledge of the trapping potential and atomic state. The model also may take into account a period of free expansion of the cloud from the trap that occurs just before the picture is snapped [1]. The experimenter then uses the fit parameters to calculate the results of interest (temperature is calculated from the Gaussian width of a gas in free expansion, for example).

This "fitting" approach to data analysis is an example of *model-based inference*. It is an appropriate and meaningful approach when the model takes into account the physics at play. The experimenter's *a priori* knowledge must be reasonably complete and accurate, and the experimental noise must not overwhelm the signal. The utility of model-based inference can rapidly disintegrate if these conditions are not met, since it can fail to extract the proper parameters of interest from data showing an unexpected effect.

The complement of a model-based approach is *model-free inference*. Model-free analyses use statistical concepts such as maximum likelihood, entropy extremization, and independence to extract the significant features buried in a collection of data. It must be emphasized that the interpretation of the features so extracted does require the experimenter to refer to some model of the experiment. "Model-free" in this case specifically refers only to the process by which the features are identified.

Powerful techniques have emerged over the past three decades and have flourished in fields where analytical models are frequently hard to develop, such as biology, economics, and the social sciences. These techniques are less familiar in physics. For experiments in which a complete model is difficult to develop, model-free inference would allow the data to speak for itself in the initial analysis, free of the interpretive bias of a wrong model.

This paper presents the application of a closely related pair of model-free analysis techniques to data derived from a Bose-Einstein condensate-based (BEC-based) Michelson interferometer realized on an atom chip. These two statistical techniques, called *principal component analysis* (PCA) and *independent component analysis* (ICA), have proven to be powerful in image- and signal-processing applications as diverse as human face recognition, brain signal analysis, economic prediction, and astronomical data processing [2–5]. While our specific focus is on interferometry, our purpose here is to use our experiment to demonstrate the utility and application of these techniques to information extraction from complex ultracold matter images in general. We believe these methods could be of use in the study of complex structures seen in gases containing vortices and gases trapped in optical lattices, for example. We demonstrate here the use of these methods to calibrate our interferometer, to identify and mitigate unforeseen sources of noise, and to uncover signal data that is partially buried in noise. Our statistical methods are able to complete these tasks in a fraction of the time needed for fitting routines to run.

## II. INTERFEROMETRY EXPERIMENT

The experiments described in this work were conducted on an upgraded version of the apparatus described in [6] and [7]. The apparatus is used to create an initially stationary condensate of $^{87}$Rb atoms in the $|F=1, m_F=-1\rangle$ ground state. The BEC is held in a microchip trap of radial frequency 80 Hz and axial frequency 4.1 Hz.

The interferometry experiment (see similar work in [6,8,9]) is initiated by the application to the BEC of a pulse of a standing wave of off-resonant light directed along the weak axis of the trap. This pulse diffracts the BEC into two packets of equal amplitude that move in opposite directions along the trap axis [see Fig. 1(a)]. The packets' initial momentum is $p_0 = \pm 2\hbar k_0$, where $k_0 = 2\pi/780$ nm [Fig. 1(b)]. The packets slow down as they propagate and eventually stop at their classical turning points after one quarter of a trap period has elapsed [Fig. 1(c)]. The packet centers are 910 µm apart at the turning point. The packets then turn around and move back towards each other, overlapping again after propagating for a total of half of a trap period, $t = \pi/\omega_z = 120$ ms [Fig. 1(d)]. The packets perform this half orbit through the trap $N$ times [repeat Fig. 1(c) and (d) until $n = N$, where $N = 2$ in most experiments]. When the packets are overlapped for the final time, the standing wave pulse is applied again [Fig. 1(e)].

Finally, the atoms are released from the trap and allowed to freely evolve for some time $\tau$, usually set to be 15 ms [Fig. 1(f)]. An absorption image, taken after the free evolution, records the momentum distribution of the atoms just before the trap was turned off. There are three peaks in the distribution, corresponding to atoms moving at $p = 0$ and $p = \pm p_0$.

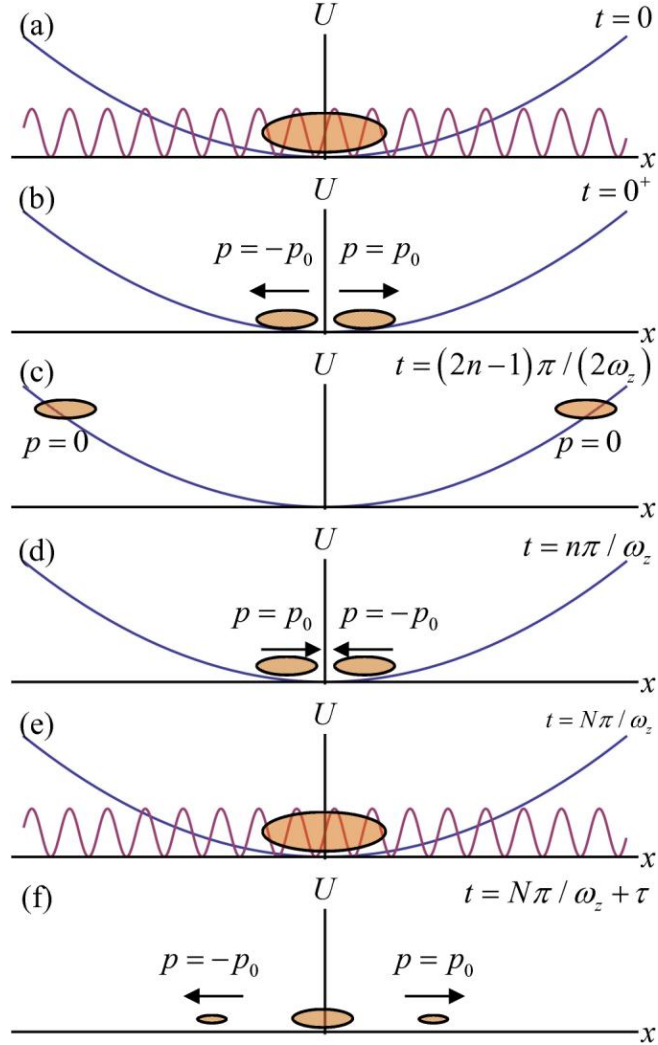

FIG. 1. (Color online) Schematic of interferometer sequence. (a) A BEC held in a cigar-shaped trap is exposed to a pulse of a standing wave of off-resonant light applied along the weak trap axis. The axial potential $U$ is plotted versus displacement $x$ from the trap bottom. (b) The BEC splits into two equal packets that propagate with opposite momentum along the weak axis. (c) The packets stop at their classical turning points, which they reach at $t = \pi / 2\omega_z$. (d) The packets overlap once more at the trap bottom at $t = \pi / \omega_z$. (c) and (d) repeat as the packets continue to propagate through the trap; in every half-period $n$ the atoms reach their maximum separation once and pass through each other once. (e) The packets continue moving through the trap for a total of $N$ half periods. The light pulse is applied to the overlapped packets. (f) The trap is shut off. The atoms are allowed to freely propagate for a time $\tau$, after which an absorption image of the momentum distribution just before trap shutdown is obtained.

The relative number of atoms in these three momentum states comprises the measurable output of the interferometer. If the packets experience no differential phase shift $\phi$ during propagation, then the second pulse will reverse the effect of the first pulse and stop all of the atoms. The momentum distribution will in this case contain a single peak centered at $p = 0$ [Fig. 2(a)]. If the packets do accumulate a nonzero $\phi$ (i.e., by the application of a magnetic field gradient or inertial force from an acceleration of the apparatus), then interference between the atoms prevents the second pulse from perfectly reversing the effect of the first pulse. Therefore only a fraction of the atoms, given by $R = \cos^2 \phi / 2$, will stop. The resultant momentum distribution images show a significant number of atoms in peaks centered at $p = \pm p_0$ [Fig. 2(b) and (c)].

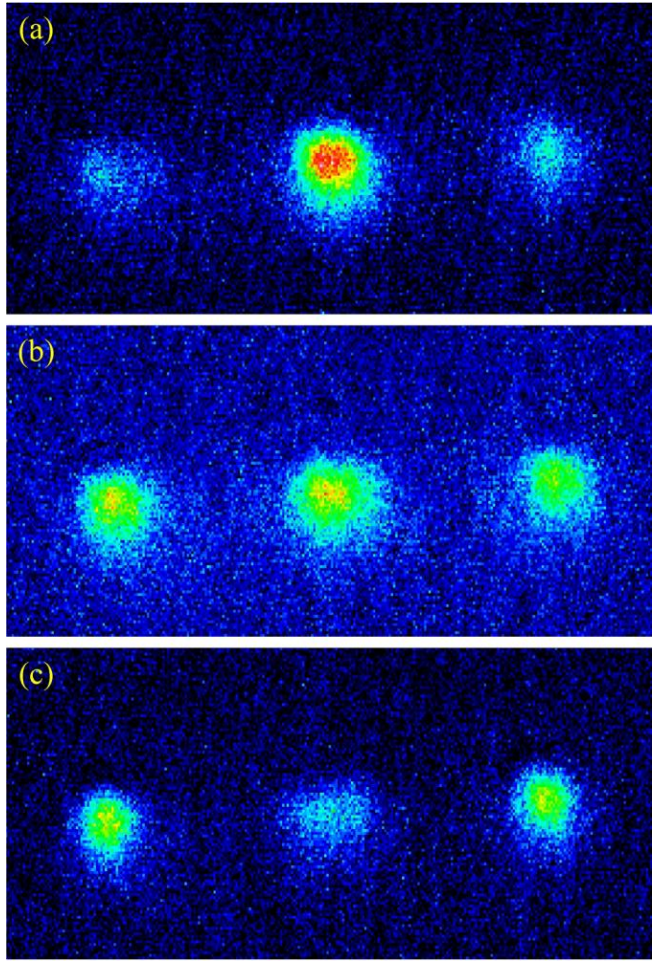

FIG. 2. (Color online) Examples of false-color images of atomic momentum distribution after an interferometer cycle for various applied phase shift $\phi$. The central cloud is centered at $p=0$, the right at $p=p_0$, and the left at $p=-p_0$. (a) $\phi \cong 2\pi m$, (b) $\phi \cong (2m+1)\pi/2$, (c) $\phi \cong 2\pi(2m+1)$ for integer $m$.

## III. MODEL-BASED IMAGE ANALYSIS

We are now presented with a problem in image analysis that must be solved before we can determine $\phi$: We must count the number of atoms in each momentum peak. Only then can we calculate the fraction $R$ in the central peak and obtain $\phi$ from

$R = \cos^2 \phi/2$. A model-based approach to this problem (which has been applied to the previous experiments of this type referenced above) consists of fitting the three peaks to Gaussian or Thomas-Fermi distributions, followed by calculating the number of atoms in each peak from the fit parameters [1].

This approach has an inherent weakness: There is no simple *a priori* model for the shape of the momentum distribution of the atoms after the interferometry cycle. The atoms begin the experiment in a pure, stationary BEC with a simple Thomas-Fermi distribution. During the experiment, the two wave packets move through each other multiple times. With each pass, a fraction of the atoms undergo collisions and are scattered out of the ballistically propagating condensate packets. Some of these ejected atoms remain in the trap for the duration of the experiment and are recorded in the absorption images. These ejected atoms are no longer coherent with the condensate packets, but a three-peak fit is unable to distinguish the "noise" atoms from the coherent atoms. The noise therefore results in an apparent decrease in the interferometry signal's visibility. Furthermore, simple routines have trouble fitting three independent shapes to the image when the space between the peaks contains a significant number of the scattered noise atoms.

Of additional concern, the fitting of three independent two-dimensional shapes to each image in a data set that contains upwards of 100 images is computationally intensive. Our fitting implementation took up to an hour to completely fit such a set of images. The slow speed prevents adjustment of experimental parameters in real time, a shortcoming that would become a serious problem in an apparatus with the rapid BEC production rate needed to take high-bandwidth measurements.

Finally, the peaks themselves are often clearly not well-described by the simple Thomas-Fermi distribution (see Fig. 3 for an example). As noted by Horikoshi and Nakagawa [9], $\phi$ is expected to be spatially inhomogeneous in experiments of this type because of dephasing effects. Such inhomogeneity causes the three peaks in the momentum distribution to acquire spatial structure. If one attempts to fit a Thomas-Fermi distribution to the results of an experiment in which the interferometer suffers from dephasing, the fit will integrate over the spatial information and suppress the interferometer's contrast. The signal would still be present in the images, but the use of the incorrect model would bury it. Other unidentified symptoms of noise or unexpected physical processes could similarly escape detection if the data were fit by a poor model.

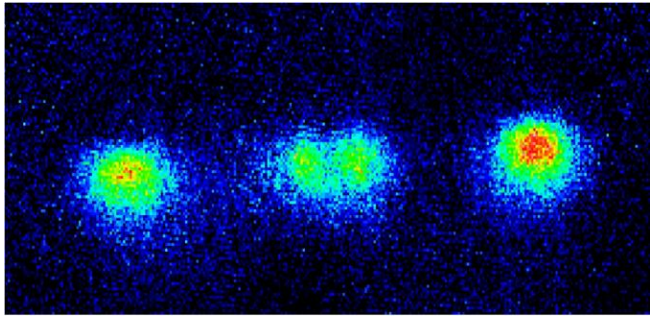

FIG. 3. (Color online) A false-color interferometer result in which spatial dephasing has occurred.

## IV. STATISTICAL IMAGE ANALYSIS

In place of fitting the images to a model, a statistical method might be used to extract and analyze the significant features. In statistical image analysis, it is useful to represent each pixilated, grayscale image as a $p$-dimensional vector, where $p$ is the number of pixels. The vector's projection along dimension $j$ is equal to the value of pixel $j$ of the image. For a very coarse camera that records only two pixels, a set of (totally independent) images could be plotted as in Fig. 4(a). An analysis of a set of images is then carried out by using the statistics of the corresponding set of vectors to find a more useful basis in which to represent the data.

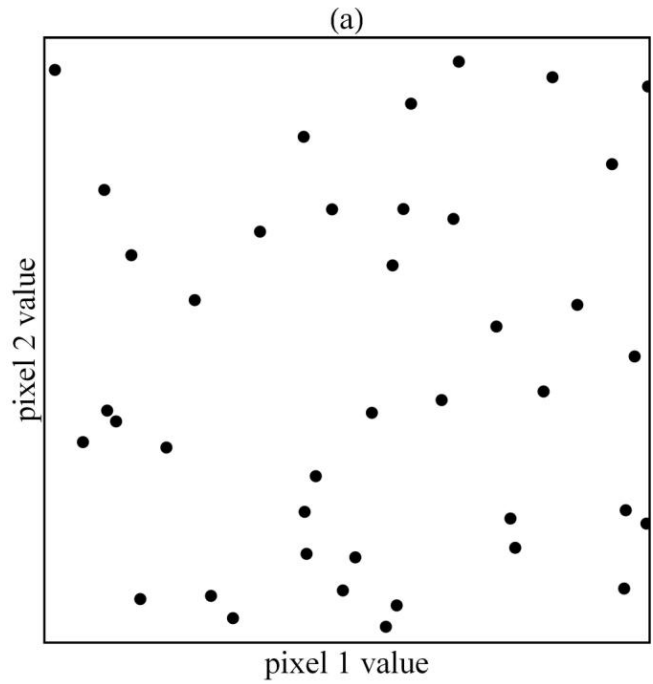

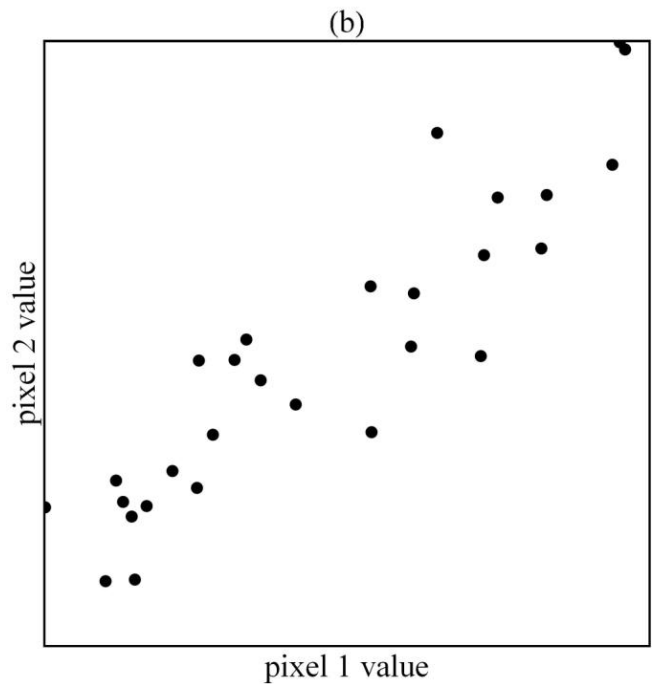

FIG. 4. Representation of coarse two-pixel images in vector form. (a) A set of independent, unrelated images. (b) A set of images in which the variation of the two pixels is not independent; this could be a set of similar images in which some detail is varied from shot to shot.

To illustrate the limits of the original pixel basis, consider the case of a set of images from our interferometry experiments. Each image contains three spatially localized features: the peaks in the momentum distribution. Each peak is recorded in the frame by the values of a localized subset of pixels. We know that the relative prominence of the central peak changes as $\phi$ is varied; this variation is recorded by a change in the value of all of the pixels in the center of the frame. In the vector representation, the components corresponding to those central pixels do not vary independently of each other. A plot of the vector representation of a two-pixel image of just the center of the central peak would look something like Fig. 4(b). It is clear that the "pixel basis" in which we have represented the image vectors is not a linearly independent choice of basis.

The image vectors may instead be represented in a different basis. A natural choice requires each basis vector to consist of a linear combination of pixels that vary independently from any other grouping of pixels, portraying a single image characteristic that varies independently of all other characteristics. The independently varying characteristic (which the experimenter can conveniently examine by representing the basis vector itself as an image) is connected to the variation of a parameter of the experiment. In a well-controlled experiment, most of the differences between the images would be quantified by the coefficient of the single basis vector corresponding to the experimentally controlled parameter. The variation of the other coefficients would represent unwanted noise processes.

PCA and ICA are both algorithms that try to find this statistically independent basis. The difference between the two, to be detailed in the next section, is that ICA uses a higher-order test of independence to find the new basis than PCA uses. An example of a subset of the basis images found by these algorithms for a given data set is shown in Fig. 5 (the basis images computed from another data set would appear slightly different). The data image is encoded by a linear combination the set's mean image (last image in the sum) and the basis images. This example illustrates how straightforward it usually is to assign a physical process to a given basis image. Note that even though the algorithms used to find the basis images are model-free, it is up to the experimenter to use a model to interpret the meaning of the basis vectors. To begin the process, we must find the basis image corresponding to the primary experimental signal, a change in the fraction of atoms found in the central momentum peak. The third basis image in Fig. 5 represents precisely that change. An image in which all of the atoms were in the center would have a large, positive value of $y_3$, for example. An examination of the other two basis images shows the experimenter the two dominant noise processes recorded in the data: The first corresponds to an overall fluctuation in the number of atoms in the experiment, and the second to a fluctuation in the vertical position of the atoms at the moment of the exposure. The experimenter may use this information to debug and reduce the experimental noise. The complete basis contains another $p-3$ basis images, but in most of our data sets their coefficients vary only slightly from image to image. These images represent detection noise.

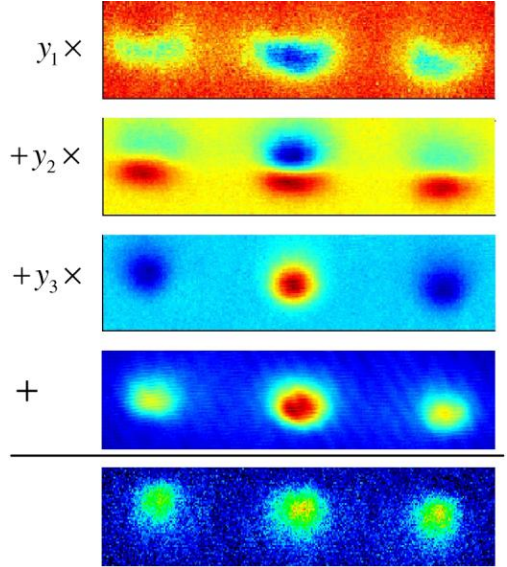

FIG. 5. (Color online) An example showing how a image from the interferometry experiment may be represented as a linear combination of independent basis images, each representing a single physical parameter in the experiment, plus the mean image of the set. False color is used to represent pixel value in all images. Varying $y_1$ changes the overall number of atoms; varying $y_2$ changes the overall vertical position of the three peaks; varying $y_3$ changes the fraction of atoms in the central peak (the primary experimental signal).

The process of recovering the signal is now straightforward. The coefficient $y_3$ is proportional to the fractional population that was previously extracted by fitting the images:

$$y_3 = A\cos\phi = A(2R-1). \qquad (1)$$

The coefficient $y_3$ now constitutes the primary measurement of the state of the interferometer in any given attempt. The phase $\phi$ (modulo $2\pi$), amplitude $A$, and estimates of the visibility of the fringes, may be extracted by fitting a set of $y_3$ values to $y_3 = A\cos\phi$ or to the expected statistical distribution based on that equation. This procedure does require fits to a model, but the fits are applied to the output of the image analysis process rather than to the raw images themselves. The distinction is analogous to the difference between using a model to explain the dependence of experimental parameters on an independent variable and using Gaussian fits in the first place to extract those parameters from a set of raw images. Parameter extraction of this type is beyond the scope of the present work, where the focus is on PCA and ICA themselves.

Figure 6 shows the result of carrying out PCA on a typical data set of over 100 images, while Fig. 7 shows the result of ICA carried out on the same images. We observe that the higher-order test used in ICA does a better job of isolating $R$ than the lower-order test of PCA does; the second principal component in Fig. 6 looks most like it describes the signal, but it is clear that varying the third principal component will slightly change $R$ as well. On the other hand, the third independent component in Fig. 7 is the only one of the three that encodes the variation of $R$ that we seek.

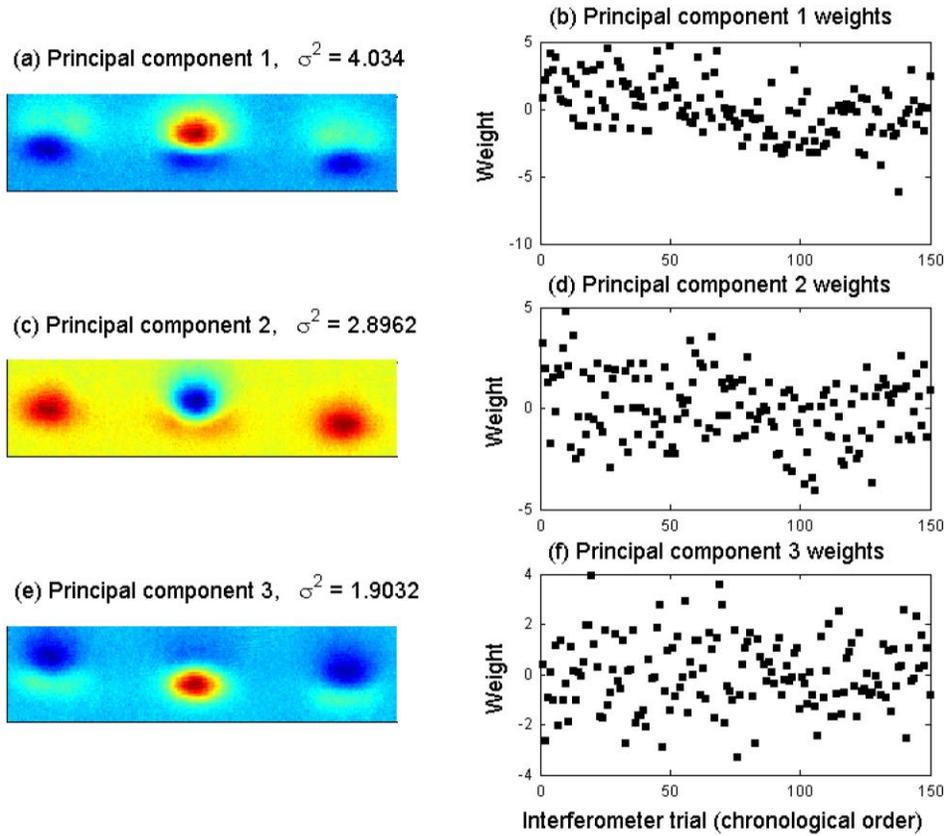

FIG. 6. (Color online) Results of PCA of a typical interferometry data set. The three principal components basis images (PCs) shown in false color account for most of the total pixel variance ($\sigma^2$) in the data set. The experimental signal $R = \cos^2 \phi/2$ can be related to PC 2, but it is clear that some of it is mixed into PC 3.

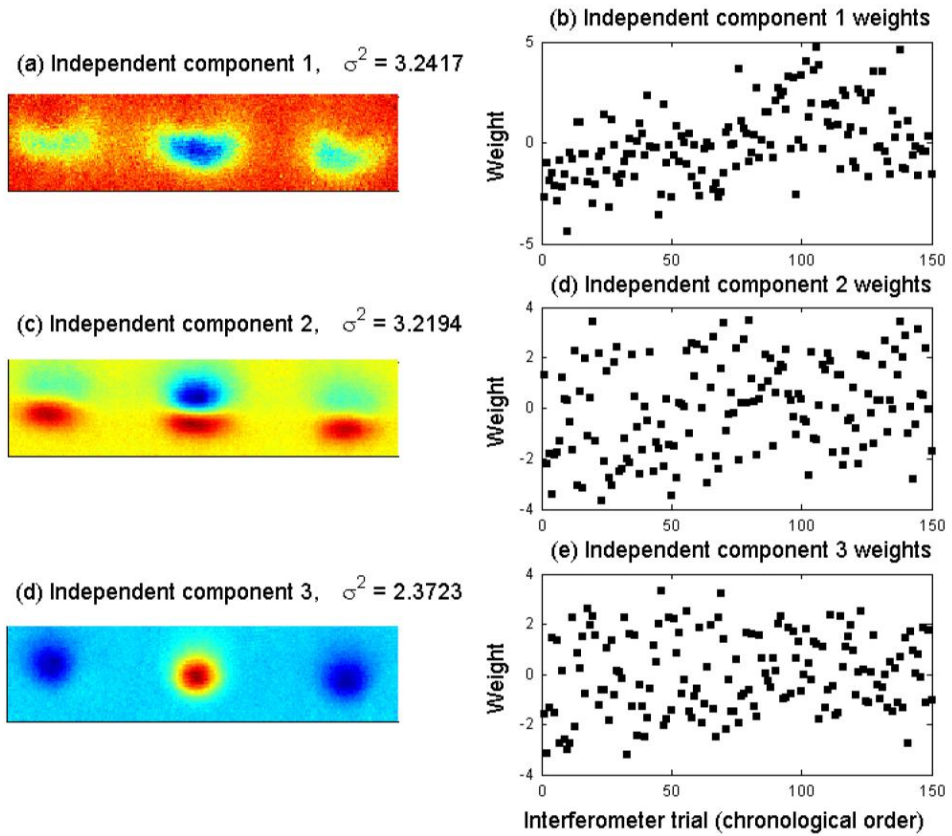

FIG. 7. (Color online) False-color results of ICA of the same data set used in Fig. 6. Independent component basis image (IC) 3 is clearly the experimental signal; it is not mixed into the other two ICs.

Figure 7 presents a typical result of our interferometry work, taken after a full trap period (240 ms) of propagation time. A qualitative inspection of the raw images from data sets of this type (see Fig. 2 for examples) shows a fairly strong shot-to-shot variance of $R$, and therefore of the phase shift $\phi$. However, in this data set, $\phi$ was not intentionally varied. Studies of the cause of the phase randomization, using the output of ICA, are ongoing [10].

We have also successfully used PCA to calibrate the apparatus before a long data run. In our experiments, the timing of the second optical pulse is a critical experimental parameter. For the atom packets to be maximally overlapped at the moment of the second pulse, the separation in time between the two pulses must be a multiple (or one-half of a multiple) of the trap period. However, it slowly drifts by hundreds of microseconds. The problem is compounded by the 90 s cycle time of the apparatus; a data set of 100 attempts represents more than 2 hours of accumulated drift. To recalibrate before taking a set, we vary this parameter around the expected value and perform ten attempts of the interferometry cycle at each value. The attempts carried out at a given pulse time are done with widely varying values of $\phi$ (applied with a magnetic field gradient) to ensure high interference contrast from shot to shot. We then carry out PCA separately on each small set of images taken at one particular pulse time. At the pulse time that results in the highest interference contrast, the variance of the coefficient of the basis image corresponding to the phase shift signal will be maximized. Figure 8 shows the result of a typical calibration run; a value of zero variance has been assigned for recombination times at which PCA did not produce a basis image that could be positively identified with the interference signal. Since PCA is computationally fast, a plot like Fig. 8 can be generated as quickly as the data are taken.

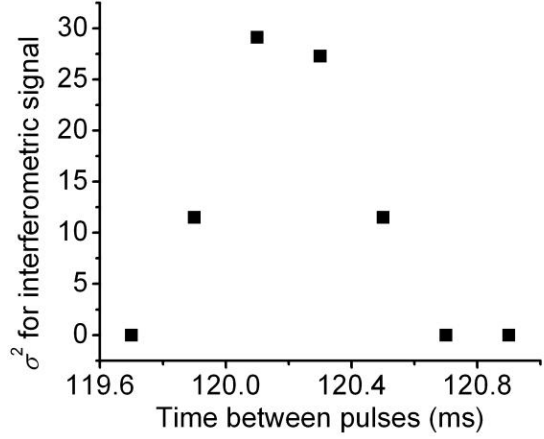

FIG. 8. Plot of the variance of the coefficient of the basis image representing the interference signal vs time between optical pulses; a value of zero variance is used to indicate times at which the interference signal cannot be identified with a single basis image.

## V. DETAILS OF ALGORITHMS

The analysis described in the previous section can be used to illustrate the distinction between the PCA and ICA algorithms. Recall that the goal is to represent an image $\mathbf{X}_i$ as a linear combination of set of basis images: $\mathbf{X}_i = \mathbf{M} + A_{1i}\mathbf{u}_1 + A_{2i}\mathbf{u}_2 + \ldots + A_{pi}\mathbf{u}_p$; here, $\mathbf{M}$ is the mean image of the set. The basis images calculated by PCA [11] are orthonormal to each other. For a set of $N$ image vectors represented in this principal component (PC) basis, the set of coefficients of any one basis image $j$ is required to be *statistically uncorrelated* with the set of coefficients of any other basis image $k$:

$$\sigma_{jk}^2 = \frac{1}{N-1}\sum_{i=1}^{N} Y_{ji}^{(P)} Y_{ki}^{(P)} = 0. \tag{2}$$

In this equation, $Y_{ji}^{(P)}$ is the coefficient of the PC basis image $j$ in the representation of image $i$. In contrast, ICA [12] applies a considerably stronger test to the statistics of the weights; the basis it calculates is, however, not necessarily orthogonal. The coefficients of the independent component (IC) basis are *statistically independent*, defined as

$$\frac{1}{N}\sum_{i=1}^{N} g\left(Y_{ji}^{I}\right) h\left(Y_{ki}^{I}\right) - \frac{1}{N^2}\left[\sum_{i=1}^{N} g\left(Y_{ji}^{I}\right)\right]\left[\sum_{i=1}^{N} h\left(Y_{ki}^{I}\right)\right] = 0. \qquad (3)$$

Here, $Y_{ji}^{(I)}$ is the coefficient of the IC basis image $j$ in the representation of image $i$, and $g$ and $h$ are any integrable functions. Two coefficients that are independent are necessarily also uncorrelated, since the definition of decorrelation can be recovered by setting $g(x) = h(x) = x$ (as long as the mean value of each coefficient is zero). Independence can therefore be thought of as a higher-order decorrelation.

One might conclude from the preceding paragraph that ICA, with its use of higher-order statistics, is always the more useful technique. It is indeed more reliable at the task of extracting the experimental signal. There exists a class of tasks, however, for which PCA remains the method of choice. Recall that for any set of vectors represented in two different orthonormal bases, the sum of the variances of the coefficients in basis 1 must be equal to the sum of the variances of the coefficients in basis 2. This rule applies to the image analysis problem as well, such that all of the image-to-image variation is preserved intact when a set of images is represented in the PC basis. Since the overall variance is conserved, the variance of a given PC coefficient can be used to rank how much of the overall variation in the data is accounted for by that PC. This ranking is essential in the timing calibration procedure described above. In contrast, the IC basis is not orthogonal. The total variance of the coefficients of the IC basis images is not equal to the total variance of the original pixels. For this reason, the IC basis images cannot be easily ranked by their variance.

## A. Principal component analysis

Having explained the difference between the two algorithms, we describe the details of the calculations. PCA is a textbook application of linear algebra [11], so the first step in the process is to represent a set of images in matrix form. Consider a set of $N$ images of $p$ pixels each. The pixel values of each image $i$ are written as the components of a vector $\mathbf{X}_i$. If the "mean image", defined as

$$\mathbf{M} = \frac{1}{N}\left[\mathbf{X}_1 + \ldots \mathbf{X}_N\right], \tag{4}$$

is subtracted from each image, a $p \times N$ matrix of the images in mean-deviation form can be constructed:

$$B = \left[\mathbf{X}_1 - \mathbf{M} \quad \mathbf{X}_2 - \mathbf{M} \quad \cdots \quad \mathbf{X}_N - \mathbf{M}\right] \\ = \left[\hat{\mathbf{X}}_1 \quad \hat{\mathbf{X}}_2 \quad \cdots \quad \hat{\mathbf{X}}_N\right]. \tag{5}$$

The $p \times p$ sample covariance matrix of the image set is calculated from $B$:

$$S = \frac{1}{N-1} BB^T. \tag{6}$$

The diagonal element $S_{jj}$ is the sample variance of pixel $j$:

$$S_{jj} = \sigma_{jj}^2 = \frac{1}{N-1} \sum_{i=1}^{N} B_{ji}^2, \tag{7}$$

and the total variance of all the pixels is tr $S$. The off-diagonal element $S_{jk}$ is the covariance of pixels $j$ and $k$:

$$S_{jk} = \sigma^2_{jk} = \frac{1}{N-1}\sum_{i=1}^{N} B_{ji}B_{ki}. \tag{8}$$

If $\sigma^2_{jk} = 0$, then pixels $j$ and $k$ are uncorrelated.

PCA uses the covariance matrix to calculate a new basis for the images. Each column $i$ of $B$ (representing one image $\hat{\mathbf{X}}_i$) is a linear combination of the original pixel basis vectors:

$$\hat{\mathbf{X}}_i = B_{1i}\begin{bmatrix}1\\0\\\vdots\\0\end{bmatrix} + B_{2i}\begin{bmatrix}0\\1\\\vdots\\0\end{bmatrix} + \ldots + B_{Ni}\begin{bmatrix}0\\0\\\vdots\\1\end{bmatrix}. \tag{9}$$

In the PCA basis, the image is represented as a linear combination of a new orthonormal set of $N$ vectors/images:

$$\hat{\mathbf{X}}_i = Y_{1i}\mathbf{u}_1 + Y_{2i}\mathbf{u}_2 + \ldots + Y_{Ni}\mathbf{u}_N. \tag{10}$$

We assume here that $N < p$, a reasonable assumption for high-resolution digital images. These basis images can be written as a $p \times N$ orthonormal matrix $P = \begin{bmatrix}\mathbf{u}_1 & \mathbf{u}_2 & \cdots & \mathbf{u}_p\end{bmatrix}$, and are referred to as the "principal components" (PCs) of the data set. The two representations of the data set are related by

$$B = PY. \tag{11}$$

Here and in Eq. (10), the matrix $Y$ is an $N \times N$ matrix in which column $i$ is the set of coefficients (or "weights") of image $i$ in the PC basis.

As defined in Eq. (2), the set of coefficients of any PC $j$ must be uncorrelated with the set of coefficients of any other PC. This requirement is met if and only if the sample covariance matrix of the PC basis,

$$D = \frac{1}{N-1} YY^T, \qquad (12)$$

is diagonal. Using Eqs. (6) and (11), along with the orthonormality of $P$, Eq. (12) may be rewritten as

$$\begin{aligned} D &= \frac{1}{N-1} P^T BB^T P \\ &= P^T SP \end{aligned} \qquad (13)$$

The sample covariance matrices of the two bases are seen to be related by

$$S = PDP^T. \qquad (14)$$

If $D$ is diagonal, then the PCs are the unit eigenvectors of $S$. The eigenvalue of a PC is the variance of its coefficient. It is often referred to as the "strength" of the PC.

It only remains to show explicitly how to reconstruct an image using the PC basis. Assume the PCs have been sorted in order of decreasing strength. Then, using Eqs. (5) and (10), image $i$ is written in the new basis as

$$\mathbf{X}_i = \mathbf{M} + Y_{1i}\mathbf{u}_1 + Y_{2i}\mathbf{u}_2 + \ldots + Y_{Ni}\mathbf{u}_N. \qquad (15)$$

In the analysis of a set of similar images, most of the variance will be captured by the first $l$ principal components, where $l$ is the smallest eigenvalue of $S$ above some cut-off value. As a result, the most important features of image $i$ can still be reconstructed if the image is represented only as

$$\mathbf{X}_i \approx \mathbf{M} + Y_{1i}\mathbf{u}_1 + Y_{2i}\mathbf{u}_2 + \ldots + Y_{li}\mathbf{u}_l. \qquad (16)$$

The dimensionality of each image is thereby reduced from $p$ pixels to $l$ PC weights. If each of these $l$ strongest PCs can be positively identified with the variation of a specific experimental parameter, then its weight in an image is proportional to the value of the associated parameter in that experimental trial.

### B. Independent component analysis

As mentioned above, the ICA algorithm [12,13] finds a different representation for the matrix $B$ in which the coefficients are statistically independent. The ICA representation is given by

$$B = P^I Y^I, \tag{17}$$

where $P^I = \begin{bmatrix} \mathbf{u}_1^I & \mathbf{u}_2^I & \cdots & \mathbf{u}_p^I \end{bmatrix}$ is a matrix of basis images referred to as independent components (ICs), and $Y^I$ is a new matrix of coefficients. The challenge is to find $P^I$ and $Y^I$ such that Eq. (3) is satisfied.

ICA can only find such a basis more effectively than PCA if the underlying experimental parameters are not normally distributed. This condition occurs because the algorithm uses tests on the higher-order statistical moments of the coefficients to determine independence. Since the higher-order moments of a normal distribution are all zero, PCA has already produced a statistically independent set; ICA can therefore do no better.

In our interferometry experiments, the interesting observable parameter is $R = \cos^2 \phi/2$. Its probability density for a uniform distribution of $\phi$ with zero mean and range greater than $\pi$ rad is shown in Fig. 9(a). This expected density is clearly not Gaussian, so ICA should find a better representation of this parameter than PCA. The results of the analysis back up this statement, as the histogram of a set of coefficients for an interference signal IC shown in Fig. 9(c) shows evidence of the expected peaks in the wings, while the histogram of PC coefficients [Fig. 9(b)] looks like a simple Gaussian. [The deviations of Fig. 9(b) from the ideal can be accounted for by the finite size of the data set and the addition of normally-distributed detection noise].

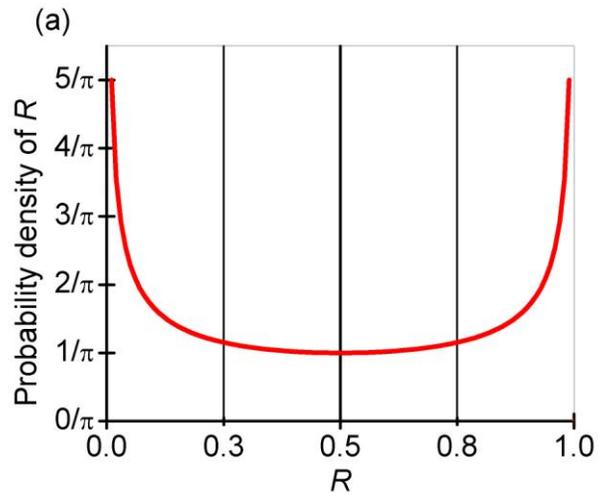

(a)

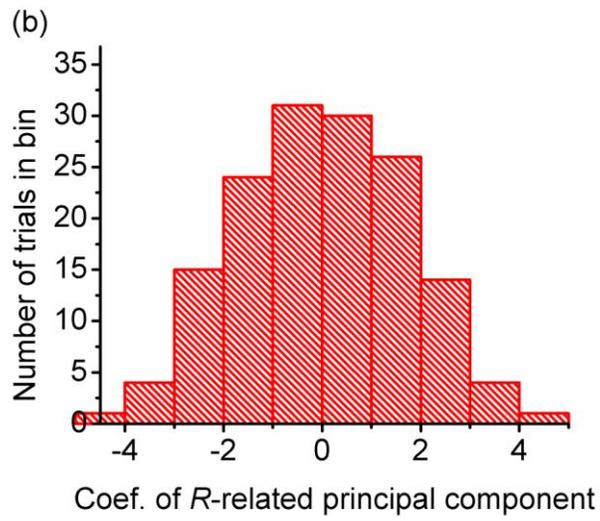

(b)

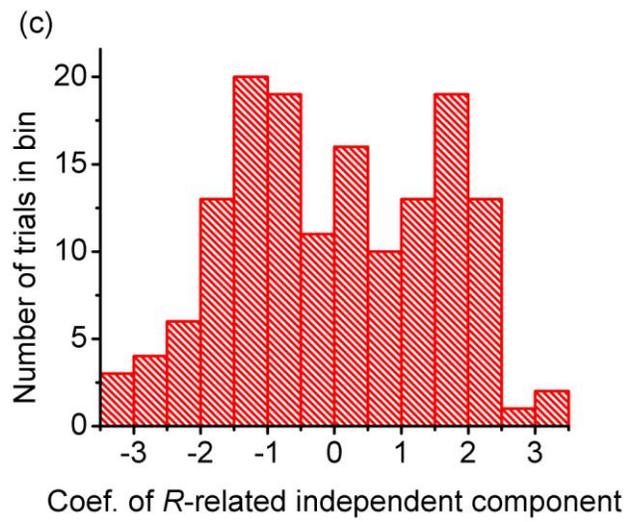

(c)

FIG. 9. (Color online) (a) The non-Gaussian probability density function of $R = \cos^2 \phi/2$ for a uniform distribution of $\phi$ with zero mean and range greater than $\pi$. (b) From a typical data set, the Gaussian distribution of the coefficient of the PC imperfectly representing the experimental signal $R$. (c) For the same data set, the clearly non-Gaussian distribution of the coefficient of the IC identified with $R$.

The relation of higher-order moments to the independence of variables is given by the central limit theorem. This states that the sum of two independent variables has a probability density that resembles a Gaussian at least as well as either of the densities of the independent variables themselves. A corollary is that the density of an independent variable is "less Gaussian" than the density of the sum of the variable and any other variable. Therefore, an ICA algorithm maximizes the "non-Gaussianity" of the density of the calculated IC's coefficients. Many possible measurements of non-Gaussianity exist; one of the simplest is the normalized kurtosis excess, defined as

$$\gamma_2 = \frac{\mu_4}{\sigma^4} - 3. \tag{18}$$

In this expression, $\sigma^4$ is the square of the variance of a variable, and $\mu_4$ is its fourth moment. Kurtosis excess measures the "peakedness" of the variable's density; a variable for which $\gamma_2 > 0$ has a density more highly peaked than a Gaussian of the same variance and vice versa. A number of more elaborate measures of non-Gaussianity exist, but all depend in some way on calculating the higher-order moments of the distributions of the coefficients.

ICA begins with a calculation of the orthonormal PC basis of a set of images. It is next assumed that each PC is actually a linear combination of the underlying ICs (which are therefore NOT orthogonal). The algorithm then forms an initial guess at the ICs by filling the columns of $P^I$ with $N$ random linear combinations of the PCs. Each column is required to be linearly independent of the other columns. The coefficient matrix of the guess is calculated by inverting $P^I$:

$$Y^I = \left[ P^I \right]^{-1} B. \tag{19}$$

A test of non-Gaussianity (perhaps kurtosis) is then applied to each row of $Y^I$. An optimization routine is used to adjust the guess of $P^I$, and the process is repeated until the coefficients of the optimized independent components are sufficiently non-Gaussian.

## VI. CONCLUSION: THE PCA/ICA TOOLBOX

In this work, we have demonstrated the utility of PCA and ICA for extracting useful data from images of a BEC experiment. PCA and ICA are methods of model-free inference, so the extracted results are not biased by any guessed-at model of the images. Together, they form a powerful toolbox for extracting results from large sets of data. PCA is useful for real-time calibration. ICA relies on a more stringent statistical test than PCA and is able to more accurately extract the phase of the interferometer. We have used these methods to examine the output of a Michelson BEC interferometer.

## ACKNOWLEDGEMENTS

The authors thank Robert Mihailovich and Ying-Ju Wang for assistance in the construction and maintenance of the apparatus used in this work. This work was supported in part by the Defense Advanced Research Projects Agency, the Army Research Office (W911NF-04-1-0043), and the National Science Foundation through a Physics Frontier Center (PHY0551010).